# WIRELESS CONGESTION CONTROL PROTOCOL FOR MULTIHOP AD HOC NETWORKS

Mahendra kumar.S
Department of Electronics and Communication Engineering
Velalar College of Engineering and Technology
Tamil Nadu, India.
.

Senthil Prakash.K
Department of Electronics and Communication Engineering
Velalar College of Engineering and Technology
Tamil Nadu, India.
.

*Abstract—* The traditional TCP congestion control mechanism encounters a number of new problems and suffers a poor performance when the IEEE 802.11 MAC protocol is used in multihop ad hoc networks. Many of the problems result from medium contention at the MAC layer. In this paper, I first illustrate that severe medium contention and congestion are intimately coupled, and TCP's congestion control algorithm becomes too coarse in its granularity, causing throughput instability and excessively long delay. Further, we illustrate TCP's severe unfairness problem due to the medium contention and the tradeoff between aggregate throughput and fairness. Then, based on the novel use of channel busyness ratio, a more accurate metric to characterize the network utilization and congestion status,

I propose a new wireless congestion control protocol (WCCP) to efficiently and fairly support the transport service in multihop ad hoc networks. In this protocol, each forwarding node along a traffic flow exercises the inter-node and intra-node fair resource allocation and determines the MAC layer feedback accordingly. The end-to-end feedback, which is ultimately determined by the bottleneck node along the flow, is carried back to the source to control its sending rate. Extensive simulations show that WCCP significantly outperforms traditional TCP in terms of channel utilization, delay, and fairness, and eliminates the starvation problem.

I. INTRODUCTION

Wireless ad hoc networks have found many applications in battlefield, disaster rescue and conventions, where fixed communications infrastructures are not available and quick network configurations are needed. To provide reliable transport services over and hence fully exploit the potential of ad hoc networks, efficient congestion control is of paramount importance. Unfortunately, the traditional TCP congestion control mechanism performs very poorly. As shown in recent studies [4].

TCP congestion control has an implicit assumption, i.e., any packet loss is due to network congestion. However, this assumption is no longer valid in the ad hoc networks as packet losses may well be due to channel errors, medium contention, and route failures.

In this paper, I mainly focus on the problems arising from medium contention when the IEEE 802.11 MAC protocol is used in the multi-hop ad hoc networks. I show that a rate based congestion control protocol is more appropriate than its window based counterpart in multihop ad hoc networks. We illustrate the close coupling between congestion and medium contention, which explains the instability of TCP. We also find that the optimum congestion window size of TCP may be less than one even in a very simple topology, say chain topology, in order to maximize the end-to-end throughput and minimize the end-to-end delay. Thus TCP tends to overshoot the network capacity and its granularity of sending rate adjustment is too coarse because the minimum increase in window size is the size of one packet upon each TCP acknowledgment or during each round trip time. Then we further show that medium contention also results in severe unfairness and starvation problems for TCP flows. Therefore, we conclude that congestion control, fairness and medium contention are all closely coupled in multihop ad hoc networks.

I propose a new wireless congestion control protocol (WCCP) based on the channel busyness ratio. In this protocol, each forwarding node determines the inter-node and intra-node fair channel resource allocation and allocates the resource to the passing flows by monitoring and possibly overwriting the feedback field of the data packets according to its measured channel busyness ratio. The feedback is then carried back to the source by the destination, which copies it from the data packet to its corresponding acknowledgment. Finally, the source adjusts the sending rate accordingly. Clearly, the sending rate of each flow is determined by the channel utilization status at the bottleneck node. In this way, WCCP is able to approach the max-min fairness in certain scenarios. We compare WCCP with TCP through extensive simulations in Section IV. We observe that WCCP significantly outperforms TCP in terms of channel utilization, delay, and fairness. Especially, it solves the starvation problem suffered by TCP.

*A. MAC layer*

The Media Access Control data communication protocol sub-layer, also known as the Medium Access Control, is a sublayer of the Data Link Layer. It provides addressing and channel access control mechanisms that make it possible for several terminals or network nodes to communicate within a multipoint network [3][5], typically a local area network or metropolitan area network . The hardware that implements the MAC is referred to as a Medium Access Controller. The MAC sub-layer acts as an interface between the Logical Link Control sub layer and the network's physical layer. The MAC layer emulates a full-duplex logical communication channel in a multipoint network. This channel may provide unicast, multicast or broadcast communication service.

*B. AODV*

Ad hoc On-Demand Distance Vector Routing is a routing protocol for mobile ad hoc networks and other wireless ad-hoc networks[2][4]. AODV is capable of both unicast and multicast routing. It is a reactive routing protocol, meaning that it establishes a route to a destination only on demand. In contrast, the most common routing protocols of the Internet are proactive, meaning they find routing paths independently of the usage of the paths. AODV is, as the name indicates, a distance-vector routing protocol. AODV avoids the counting-to-infinity problem of other distance-vector protocols by using sequence numbers on route updates, a technique pioneered by DSDV.

## II. LITERATURE SURVEY

Good and bad network performance is largely dependent on the effective implementation of network protocols. TCP, easily the most widely used protocol in the transport layer on the Internet (e.g. HTTP, TELNET, and SMTP), plays an integral role in determining overall network performance. Amazingly, TCP has changed very little since its initial design in the early 1980's. A few "tweaks" and "knobs" have been added, but for the most part, the protocol has withstood the test of time. However, there are still a number of performance problems on the Internet and fine tuning TCP software continues to be an area of work for a number of people. The design of TCP was heavily influenced by the end-to end argument.  Method of handling congestion and network overload. Are the key components of the end-to-end argument.

The four congestion control algorithms used in TCP are, Slow Start, Congestion Avoidance, Fast Retransmit and Fast Recovery. Slow Start, a requirement for TCP software implementations is a mechanism used by the sender to control the transmission rate [3], otherwise known as sender-based flow control. This is accomplished through the return rate of acknowledgements from the receiver. In other words, the rate of acknowledgements returned by the receiver determines the rate at which the sender can transmit data. When a TCP connection first begins, the Slow Start algorithm initializes a congestion window to one segment, which is the maximum segment size (MSS) initialized by the receiver during the connection establishment phase. When acknowledgements are returned by the receiver, the congestion window increases by one segment for each acknowledgement returned. Thus, the sender can transmit the minimum of the congestion window and the advertised window of the receiver, which is simply called the transmission window.

In the Congestion Avoidance [3] [9] algorithm a retransmission timer expiring or the reception of duplicate ACKs can implicitly signal the sender that a network congestion situation is occurring. The sender immediately sets its transmission window to one half of the current window size (the minimum of the congestion window and the receiver's advertised window size), but to at least two segments. If congestion was indicated by a timeout, the congestion window is reset to one segment, which automatically puts the sender into Slow Start mode. If congestion was indicated by duplicate ACKs, the Fast Retransmit and Fast Recovery algorithms [3] are invoked. As data is received during Congestion Avoidance, the congestion window is increased. However, Slow Start is only used up to the halfway point where congestion originally occurred. This halfway point was recorded earlier as the new transmission window. After this halfway point, the congestion window is increased by one segment for all segments in the transmission window that are acknowledged. This mechanism will force the sender to more slowly grow its transmission rate, as it will approach the point where congestion had previously been detected.

The Media Access Control data communication protocol sub-layer, also known as the Medium Access Control, is a sub layer of the data link layer[3]. It provides addressing and channel access control mechanisms that make it possible for several terminals or network nodes to communicate within a multipoint network, typically a local area network or metropolitan area network. The MAC sub-layer acts as an interface between the Logical Link Control sub layer and the network's physical layer. The MAC layer emulates a full-duplex logical communication channel in a multipoint network. This channel may provide unicast, multicast or broadcast communication service.

The channel access control mechanisms provided by the MAC layer are also known as a multiple access protocol. The most widespread multiple access protocol is the contention based CSMA/CD protocol used in Ethernet networks. This mechanism is only utilized within a network collision domain, for example an Ethernet bus network or a hub network. An Ethernet network may be divided into several collision domains, interconnected by bridges and switches [3][5]. A multiple access protocol is not required in a switched full-duplex network, such as today's switched Ethernet networks, but is often available in the equipment for compatibility reasons.

Several works have pointed out that greedy TCP can result in severe congestion in ad hoc networks and hence performance degradation. RED is one of the schemes proposed to overcome this problem [2]. Random early detection, also known as random early discard or random early drop is an active queue management algorithm. It is also a congestion avoidance algorithm [6].

In the traditional tail drop algorithm, a router or other network component buffers as many packets as it can, and simply drops the ones it cannot buffer. If buffers are constantly full, the network is congested. Tail drop distributes buffer space unfairly among traffic flows. Tail drop can also lead to TCP global synchronization as all TCP connections "hold back" simultaneously, and then step forward simultaneously. Networks become under-utilized and flooded by turns. RED addresses these issues. It monitors the average queue size and drops (or marks when used in conjunction with ECN) packets based on statistical probabilities. If the buffer is almost empty, all incoming packets are accepted. As the queue grows, the probability for dropping an incoming packet grows too. When the buffer is full, the probability has reached 1 and all incoming packets are dropped. RED is considered more fair than tail drop [2].

The more a host transmits, the more likely it is that its packets are dropped. Early detection helps avoid global synchronization. RED makes Quality of Service differentiation impossible. Weighted RED [5] and RED In/Out provide early detection with some QoS considerations. The Adaptive RED algorithm infers whether to make RED more or less aggressive based on the observation of the average queue length. If the average queue length oscillates around min threshold then Early Detection is too aggressive. On the other hand if the average queue length oscillates around max threshold then Early Detection is being too conservative. The algorithm changes the probability according to how aggressive it senses it has been discarding traffic.

RED has some disadvantages too, Unfairness is caused as nodes drop packets unaware of other nodes and Queue at any single node cannot reflect the network congestion state to overcome these problems Neighborhood Random Early Detection is proposed. It is based on the principle that Queue size of a neighborhood reflects the degree of local network congestion.

### III. SYSTEM DESIGN - WIRELESS CONGESTION CONTROL PROTOCOL (WCCP)

TCP's congestion control suffers from a coarse granularity when applied to the multihop ad hoc environment. To overcome this problem a rate based wireless congestion control protocol is defined. There are two components in WCCP. One is at the transport layer. It replaces the window adjusting algorithm of TCP with a rate control algorithm to regulate the sending rate. The other is between the networking layer and the MAC layer. It monitors and possibly modifies the feedback field in TCP data packets when it passes the outgoing packets from the networking layer to the MAC layer and the incoming packets in the reverse direction.

#### A. System Architecture

Based on the estimate of the available bandwidth, the inter-node and intra-node resource allocation schemes are proposed to determine the available channel resource for each node and for each flow passing through that node and accordingly modify the MAC layer feedback. Then an end to-end rate control scheme is proposed to carry the feedback from the bottleneck node to the source node which accordingly adjust the sending rate to make full and fair utilization of the channel resource at the bottleneck node without causing severe medium contention and packet collision.

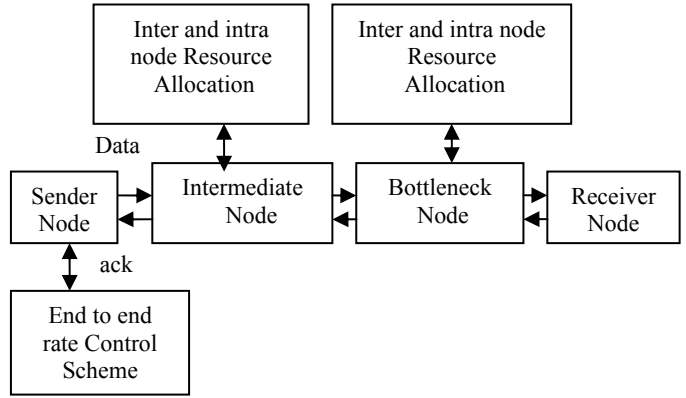

Figure 1. WCCP Mechanism

#### B. Measurement of Channel Busyness Ratio

The channel busyness ratio rb is an easily measured metric at the location of each node under the current architecture of the IEEE 802.11 standard. Notice that the IEEE 802.11 is a CSMA-based MAC protocol, working on the physical and virtual carrier sensing mechanisms. There is already a function to determine whether the channel is busy or not, i.e., the channel is determined busy when the measuring node is transmitting, receiving, or its network allocation vector indicates the channel is busy, and is idle otherwise. The channel busyness ratio can be simply calculated as the ratio of the total lengths of busy periods to the total time during a time interval, which is the average control interval.

#### C. Bandwidth Calculation

In the rate-based congestion control algorithm, to calculate the ideal sending rate, the source is in direct need of a timely and easily measured metric which should satisfy two requirements. First, since MAC contention is tightly coupled with congestion, a candidate of congestion signal should reflect the condition of MAC contention and collision. Second, in order to fully utilize the shared channel without causing severe congestion and packet collision, the candidate should indicate the available bandwidth. Channel busyness ratio rb, meets these two requirements. the channel utilization cu indicates the ratio of the channel occupancy time for successful transmissions to the total time, the normalized throughput s indicates the achievable data rate for payload divided by the channel data rate and is proportional to cu and the collision probability p indicates the probability that each transmission encounters a collision. the threshold thb should be chosen such that

$$rb \approx cu(rb \leq thb)$$

Since the performance of rb is not sensitive to n, we can fix n and observe the effect of the payload size. After choosing thb, according to Equation (3), we can estimate the available bandwidth of each node, denoted by BWa, as follows

$$BWa = BW (thb − rb) data/Ts, (rb<thb) \quad (1)$$

Where BW is the transmission rate in bits/s for the DATA packets, data is the average payload size in unit of channel occupancy time, and Ts is the average time of a successful transmission at the MAC layer.

*D. Inter Node Resource Allocation*

According to equation above equation, each node could calculate the total available bandwidth for its neighborhood based on the measured channel busyness ratio in a period called average control interval, denoted by Tc. To determine the available bandwidth for each node, WCCP accommodates the channel resource ΔS for each node proportionally to its current traffic load S in Tc.

$$\Delta S = (thb − rb)/rb * s \quad (2)$$

Because both the incoming traffic and outgoing traffic of each node consume the shared channel resource, S should include the total traffic (in bytes), i.e., the sum of the total incoming and outgoing traffic. There is two cases when we compare the observed rb with thb, i.e., rb <thb and rb ≥ thb. When rb < thb , ΔS is positive, meaning we should increase the traffic Since the available bandwidth is proportional to thb − rb according to equation (1), we may increase S by such an amount that after the increase ΔS, S is proportional to thb, which is the optimal channel utilization. Actually, equation (2) achieves our desired increase as it can be easily seen that

$$(\Delta S + S)/thb=s/rb \quad (3)$$

Therefore, rb will approach thb after one average control interval Tc when all the nodes in the neighborhood increase the total traffic rate according to equation (2). When rb ≥ thb, ΔS is negative, meaning we decrease the traffic. In this case, however, the linear relationship between the available bandwidth and rb no longer exists, and the collision probability increases dramatically as the total traffic rate increases. In addition, when the node enters saturation, both collision probability and rb amount to their maximum values and do not change as the traffic increases, although the total throughput decreases. It thus appears that ideally, WCCP needs to aggressively decrease the total traffic rate. However, since it is difficult to derive a simple relationship between the traffic rate and rb when rb ≥ thb, WCCP uses the same linear function as for the case rb < thb.. Indeed, this brings two advantages. First, as the increase and decrease use the same law, it is simple to implement at each node. Second, opting out of aggressive decrease helps achieve smaller oscillation in channel utilization.

*E. Intra Node Resource Allocation*

After calculating ΔS, the change in the total traffic or the aggregate feedback at each node, WCCP needs to apportion it to individual flows traversing that node in order to achieve both efficiency and fairness. WCCP relies on an Additive-Increase Multiplicative- Decrease policy to converge to efficiency and fairness: If ΔS > 0, all flows increase the same amount of throughput. And if ΔS < 0, each flow decreases the throughput proportionally to its current throughput.

Before determining the feedback when ΔS >= 0, WCCP needs to estimate the number of flows passing through the considered node. Again, since the channel is shared by both incoming and outgoing traffic,

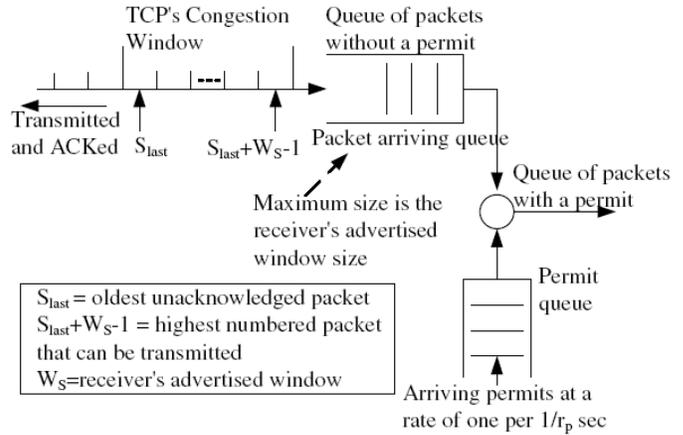

Figure 2. Rate control mechanism

The number of flows J used by WCCP should be different from the real number of flows. For those flows that either originate or terminate at the node, the node counts each as one flow, whereas for those flows only passing through the node, the node counts each as two flows, i.e., one in and one out. This is because that flows passing through the node occupy twice channel resource of that for flows originating or terminating at the node Let rpk denote the packet sending rate (pkt/s) of the flow which the kth observed packet during the period Tc at node i belongs to. Since rpkTc is equal to the number of packets which are observed from the corresponding flow at node I during Tc, we can convert the summation over each flow to the summation over each packet by the fact that, for all the packets of each flow, the summation of 1/rpkTc is equal to 1. Thus J can be calculated at node i as

$$J =1/rph*Tc \quad for\ k->1\ to\ K \quad (4)$$

Where K is the total number of packets seen by node i in Tc. Thus each node only needs to do the summation for each received and transmitted packet. Therefore, if ΔS >= 0, the increasing amount of traffic rate for each flow Cp, and per packet feedback pfk will be

$$Cp=\Delta S/Tcj \quad (5)$$

$$Pfk=Cn*Rpk/Rpk*Tc \quad (6)$$

where Cn is a constant and WCCP aims to make full use of the channel resource while not introducing severe medium contention, i.e., rb should be as close to thb as possible but never exceed thb too much. Therefore, when the aggregate feedback of previously passing packets is equal to ΔS, the node sets local feedback value as zero until the next control interval starts. With this mechanism in place, the channel busyness ratio rb should be around thb at the bottleneck nodes and be smaller at other nodes.

*F. End-To-End Rate-Based Congestion Control Scheme*

A leaky bucket (permit queue) is attached to the transport layer to control the sending rate of a WCCP sender. The permit arrival rate rp of the leaky bucket is dynamically adjusted according to the explicit feedback fb carried in the returned ACK whenever a new ACK arrives (henceforth, ACKs refer to the transport layer acknowledgments). Namely,

$$Rp = rp + fb \qquad (7)$$

To enable this feedback mechanism, each WCCP packet carries a congestion header including three fields, i.e., rp, Tc, and fb, which is used to communicate a flow's state to the intermediate nodes and the feedback from the intermediate nodes to the source. The field rp is the sender's current permit arrival rate, and the field Tc is the sender's currently used control interval. They are filled in by the sender and never modified in transit. The last field, fb, is initiated by the sender and all the intermediate nodes along the path may modify it to directly control the packet sending rate of the source.

## IV. SYSTEM IMPLEMENTATION

*A. Tool Description*

The network simulator ns-2 is an object–oriented, discrete event-driven network simulator developed at the OC Berkley and ISC ISI as part of the VINT project [VIN03]. It is a very useful tool for conducting network simulations involving local and wide area networks. In the recent years its functionality has grown to include wireless and ad hoc networks as well.

The ns-2 network simulator has gained an enormous popularity among participants of the research community, mainly because of its simplicity and modularity. The network simulation allows simulation scripts, also called simulation scenarios, to be easy written in a script-like programming language TCL. Emulation refers to the ability to introduce the simulator into a live network. Special objects within the simulator are capable of introducing live traffic into the simulator and injecting traffic from the simulator into the live network. The interface between the simulator and live network is provided by a collection of objects including Tap Agents and Network Objects. Tap agents embed live network data into simulated packets and vice-versa. Network objects are installed in tap agents and provide an entry point for the sending and receipt of live data. Both objects are described in the following sections. When using the emulation mode, a special version of the system scheduler is used: the Real Time Scheduler. This scheduler uses the same underlying structure as the standard calendar-queue based scheduler, but ties the execution of events to real time

More complex functionality relies on C++ code that either comes with ns-2 or is supplied by the user. The utilization of the two programming languages increases the flexibility of ns-2. C++ is mainly used for event handling and per-packet processing tasks for which TCL would become too slow. TCL is most commonly used for simpler routing protocols, general ns-2 code and simulation scenario scripts. The usage of TCL for simulation scenario scripts allows the user to change parameters of a simulation without having to recompile any source code.

Simulations in ns-2 can be logged to trace files, which include detailed information about packets in the simulation and allows for post-run processing with some analysis tool. It is also possible to let ns-2 generate a special trace file that can be used by NAM (Network Animator), a visualization tool that is part of the simulator distribution.

*B. Resource Requirements*

- Hardware requirements

| Processor Type | : | Pentium-IV,512MB RAM |
|---|---|---|
| Hard Disk | : | 20 GB |

- Software requirements

| Operating System | : | Linux |
|---|---|---|
| Programming Language | : | C++ |
| Tool | : | Network Simulator 2 |

## V. RESULTS AND DISCUSSION

*A. NAM Output*

The nam output shows the 9-node chain topology. There are three cases considered to compare the performance of TCP and Wireless congestion Control Protocol (WCCP).In case 1 there is only one flow between node0 to node 8 and in case 2 there are three flows one flow is between node0 to node4 and the second flow is between node0 to node8 and third flow is between node4 to node8.

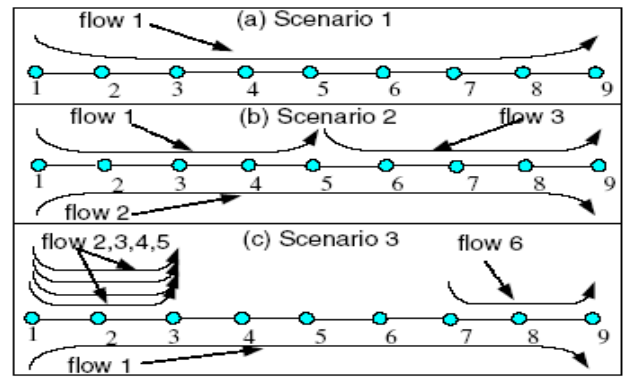

Figure 3. 9-Node Chain Topology with Different Traffic Distribution

Figure 3 shows various flows for the nine node topology. Scenario 1 has only one flow from node 1 to node9 and Scenario 2 has 3 flows and Scenario 3 has 6flows

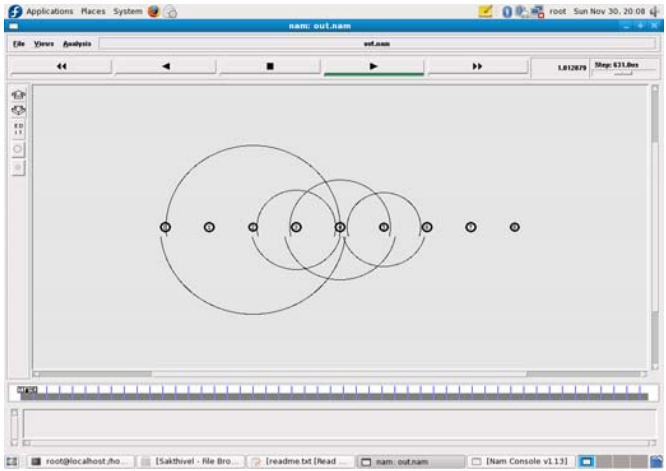

Figure 4. Nam Output

Figure 4 shows the Nam Output for the nine node topology as shown in the figure 3 it shows the Nam output for 3 flows

### B. Graphs

*1) Scenario1*

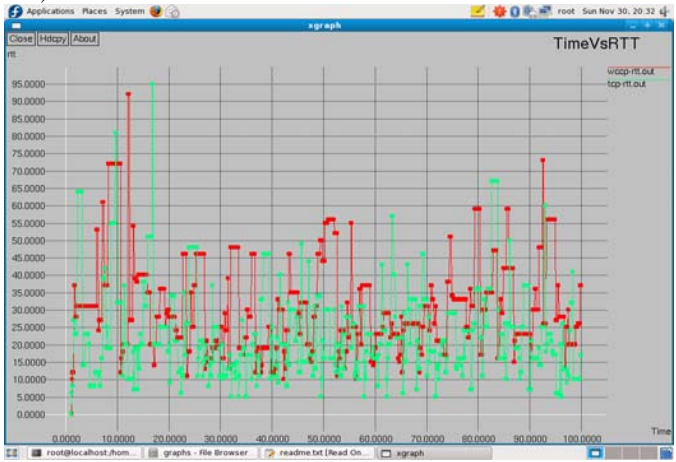

Figure 5. Round Trip Time of TCP vs WCCP

Figure 5. Shows the comparison of round trip time of TCP vs WCCP for scenario1

*2) Scenario 2*

The graph shows the comparison of throughput for TCP and WCCP. Throughput of WCCP is higher than TCP.

Figure 6 shows the throughput of TCP for various flows. Here there are three flows. Throughput of the second flow which is the main flow is much smaller than the other flows.

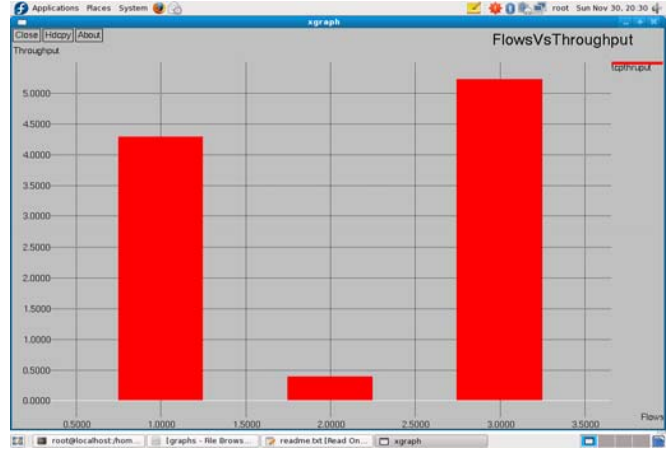

Figure 6. TCP Throughput

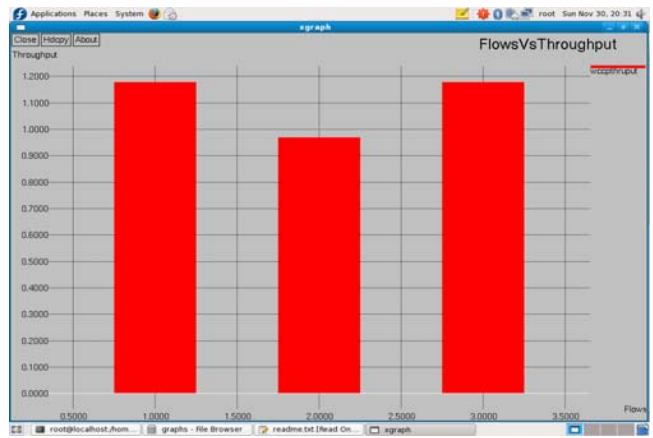

Figure 7. WCCP Throughput

Figure 7 shows the throughput of WCCP for various flows. Here there are three flows. Throughput for the second flow is very small for TCP but it is overcome by Wireless Congestion Control Protocol. Here it overcomes the problem by allocating the resources properly at each node and by providing the required feedback to the source node.

*3) Scenario 3*

The graph shows the comparison of throughput for TCP and WCCP. Throughput of WCCP is higher than TCP.

Figure 8 shows the throughput of TCP for various flows. Here there are six flows. Four flows from node1 to node3 and one from node7 to node9 and the main flow is from node1 to node9 and due to many flows the throughput for main flow is zero

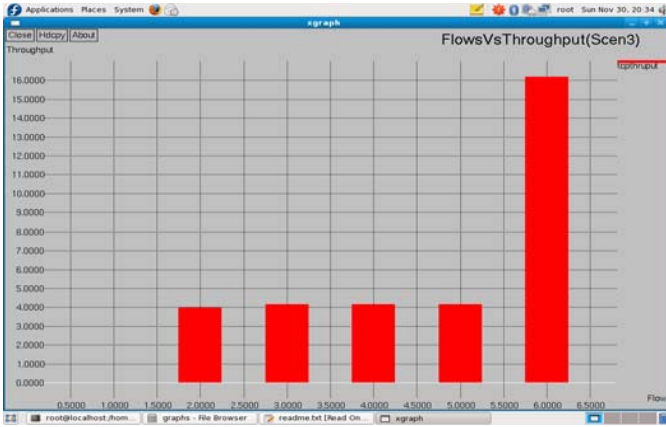

Figure 8. TCP Throughput

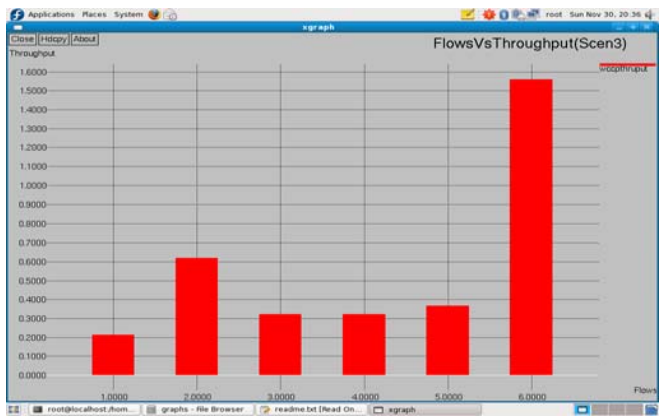

Figure 9. WCCP Throughput

Figure 9 shows the throughput of TCP for various flows. Here there are six flows. Four flows from node1 to node3 and one from node7 to node9 and the main flow is from node1 to node9 and due to many flows the throughput for main flow is zero in TCP but it is overcome by WCCP.

## VI. CONCLUSIONS AND FUTURE ENHANCEMENT

Congestion control is critical to reliable transport service in wireless multihop ad hoc networks. Unfortunately, traditional TCP suffers severe performance degradation and unfairness. Realizing that the main reason is the poor interaction between traditional TCP and the MAC layer, we propose a systematic solution named Wireless Congestion Control Protocol (WCCP) to address this problem in both layers. WCCP uses channel busyness ratio to allocate the shared resource and accordingly adjusts the sender's rate so that the channel capacity can be fully utilized and fairness is improved. We evaluate WCCP in comparison with TCP in various scenarios. The results show that our scheme outperforms traditional TCP in terms of channel utilization, end-to-end delay, and fairness, and solves the starvation problem of TCP flows.

AUTHORS PROFILE

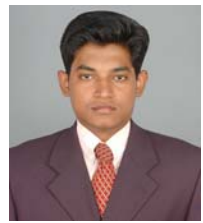

Mahendra Kumar.S received the Bachelors degree in Electronics and Communication Engineering from from Anna University, Chennai in 2006. and the Master degree in Communication System from Kumaraguru College of Technology, Coimbatore in 2009. He is currently working in Velalar College of Engineering and Technology as Lecturer of Electronics and Communication Department since 2009. His fields of interests include Ad Hoc & Sensor Network, He has published two papers in National Conferences and one International Conference in Ad Hoc Networks. He is a life member of ISTE.

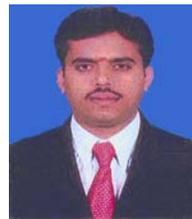

Senthil Prakash.K received the Bachelors degree in Electronics and Communication Engineering from Bharathidasan University, Tiruchirappalli in 2003. and the Master degree in Communication System from Kumaraguru College of Technology, Coimbatore in 2009. He is currently working in Velalar College of Engineering and Technology as Lecturer of Electronics and Communication Department since 2009. His area of interest includes Mobile communication and Ad Hoc Network, He has published two papers in National Conferences in Ad Hoc Networks. He is a life member of ISTE.